\documentclass[conference]{IEEEtran}
\IEEEoverridecommandlockouts
\usepackage{textcomp}
\usepackage{amsthm}
\usepackage[english]{babel}

\usepackage{graphics}
\usepackage{caption}

\usepackage{subcaption}

\usepackage{cite}
\usepackage{amsmath,amssymb,amsfonts}
\usepackage{algorithmic}
\usepackage{graphicx}
\usepackage{textcomp}
\usepackage{xcolor}
\usepackage{multicol}
\usepackage{multirow}
\usepackage{makecell}
\usepackage{array}
\newcolumntype{P}[1]{>{\centering\arraybackslash}p{#1}}
\newcolumntype{M}[1]{>{\centering\arraybackslash}m{#1}}

\def\BibTeX{{\rm B\kern-.05em{\sc i\kern-.025em b}\kern-.08em
    T\kern-.1667em\lower.7ex\hbox{E}\kern-.125emX}}
    
\begin{document}

\title{Power System Recovery Coordinated with (Non-)Black-Start Generators\\
}

\author{\IEEEauthorblockN{Meng Zhao, Patrick R. Maloney, Xinda Ke, Juan Carlos Bedoya Ceballos, Xiaoyuan Fan, Marcelo A. Elizondo }
\IEEEauthorblockA{
(meng.zhao; patrick.maloney; xke; juan.bedoya; xiaoyuan.fan; Marcelo.Elizondo) @pnnl.gov \\
\textit{Pacific Northwest National Laboratory}\\
Richland, WA}
\thanks{ This study was conducted at Pacific Northwest National Laboratory, which is operated for the U.S. Department of Energy (DOE) by Battelle Memorial Institute under Contract DE-AC05-76RL01830.}
\thanks{The authors would like to thank the Federal Emergency Management Agency (FEMA) and the Department of Homeland Security (DHS) for funding this study.}
\thanks{Corresponding author: patrick.maloney@pnnl.gov.}
}

\maketitle

\begin{abstract}
Power restoration is an urgent task after a black-out, and recovery efficiency is critical when quantifying system resilience. Multiple elements should be considered to restore the power system quickly and safely. This paper proposes a recovery model to solve a direct-current optimal power flow (DCOPF) based on mixed-integer linear programming (MILP). Since most of the generators cannot start independently, the interaction between black-start (BS) and non-black-start (NBS) generators must be modeled appropriately. The energization status of the NBS is coordinated with the recovery status of transmission lines, and both of them are modeled as binary variables. Also, only after an NBS unit receives cranking power through connected transmission lines, will it be allowed to participate in the following system dispatch. The amount of cranking power is estimated as a fixed proportion of the maximum generation capacity. The proposed model is validated on several test systems, as well as a $1393$-bus representation system of the Puerto Rican electric power grid. Test results demonstrate how the recovery of NBS units and damaged transmission lines can be optimized, resulting in an efficient and well-coordinated recovery procedure.
\end{abstract}


\section{Introduction}
\IEEEPARstart{R}{ecent} frequent black-outs demonstrate that existing power grids are vulnerable to natural disasters. For example, in 2017, significant damage was caused on the transmission and distribution systems of Puerto Rico by Hurricanes Irma and Maria, leading to one of the longest blackouts in U.S. history and leaving residents in some parts of the territory without electricity for almost a year. Moreover, this blackout also impacted other infrastructure services, e.g., communication, water and wastewater, transportation, healthcare, and critical manufacturing sectors. Therefore, it is of critical importance, that we develop tools and procedures, which can quickly restore power following an outage.

Recovering power back from an outage in a short time means to improve the system resilience.
As defined in \cite{resilience}, resilience is ``the ability to withstand and reduce the magnitude and/or duration of disruptive events, which includes the capability to anticipate, absorb, adapt to, and/or rapidly recover from such event." Multiple steps should be taken during the recovery process. Among them, one of the critical elements is effective utilization of black-start (BS) generators \cite{1995AFF}. 
BS generators, unlike non-black-start (NBS) generators, do not require an external power source (often referred to as cranking power) to begin producing power following a shutdown. Therefore, proper coordination of these resources as well as line recovery must be carefully conducted to define procedures where lines between BS and NBS units are recovered, so that NBS units can receive cranking power from BS units and begin participating in system recovery.
A typical BS operation includes three phases: system preparation (analyze system status and develop a strategy to black start the system), system restoration (BS generators are started to energize the transmission lines and other NBS units), and load restoration (re-energize more loads with more restored generators) \cite{ElectricGB}. 

Existing research considering BS units in system recovery can be largely divided into two categories: BS-based recovery optimization model; BS-targeted restoration optimization model. In the BS-based recovery model, new constraints considering BS resources are included and a corresponding optimization method is proposed. In \cite{Ding2022ASB}, a sequential BS restoration model which includes a novel topology constraint is proposed by using a mixed-integer second-order cone programming. The authors in \cite{Shi2022ATS} establish a two-stage stochastic distribution network restoration model considering BS resources and uncertainties from distributed renewable generators, in which the first stage is to solve the line recovery, and the second stage includes the scenario-wise operation of distributed generators and dispatchable loads. In \cite{Zhu2021ATD}, a dynamic programming method and pruning algorithm are utilized to solve a multi-objective optimization problem integrated with BS resources. This includes minimizing the switching time, maximizing the recovery efficiency as well as the priority recovery of critical loads.

In the BS-targeted restoration model, the goal is to optimally allocate the BS units to reduce the relevant costs. In \cite{2016OptimalBS}, the BS resource procurement decision is integrated with the restoration planning model to minimize the procurement cost. \cite{2018OptimalBS} develops a more detailed optimization model which optimally allocates BS units and simultaneously optimizes the restoration sequence. An energization binary variable for each time step and another binary variable for BS allocation are used to better coordinate these two optimizations. 

Although current power recovery research based on BS resources have achieved success in some extent, there are still some limitations: due to the additional costs required to install BS units, most research focuses on minimizing installation costs or optimized recovery coordinated with BS. However, the NBS units are ignored which can supply power to the system after being energized and can cost less than BS resources. Moreover, the cranking power needed to energize the NBS units is relatively small in comparison to other standard system loads. In this case, coordination between the NBS units and recovery process may decrease the costs as well as improve efficiency. 

To overcome the limitations existing in current research, this paper proposes a novel optimization model based on mixed-integer linear programming (MILP) to better coordinate the NBS units with the recovery process. Specifically, the features of this recovery model include:
\begin{itemize}
    \item modeling NBS units as loads before they receive cranking power and dispatchable generators afterwards;
    \item estimating cranking power as a fixed percentage of the maximum generation capacity of an NBS unit (summarized from \cite{cranking});
    \item coordination between the NBS energization and transmission line recovery.
\end{itemize}

The rest of this paper is organized as follows. Section \uppercase\expandafter{\romannumeral2} introduces the mathematical model of the recovery process. The realization of the NBS recovery is illustrated in Section \uppercase\expandafter{\romannumeral3}. Section \uppercase\expandafter{\romannumeral4} shows the simulation results on some test systems and a real power system in Puerto Rico. The conclusions are summarized in Section \uppercase\expandafter{\romannumeral5}.

\section{Mathematical Model}

Suppose a power system network $\mathcal{P}=(B, A)$, in which $B$ and $A$ represent the set of buses and lines, and $G$ is the set of generators. The established recovery model is to minimize the unserved load while satisfying the general operation constraints and proposed recovery constraints.

Within the recovery period $T$, at time step $t$, the objective function can be formulated as,

\begin{equation}
\label{obj}
    \min \sum_{t=1}^{T} \sum_{b\in B} LS_{b,t}
\end{equation}
where $LS_{b,t}$ represents the unserved load of bus $b$ at recovery step $t$.

\subsection{Operation Constraints}

In this optimization model, Kirchhoff's Current Law (KCL) is considered for the transmission and generation models \cite{2018WindCG},

    \begin{equation}
    \label{ope}
    \begin{split}
       \sum_{g_b\in G} P_{g_b,t} + LS_{b,t} & + \sum_{a(b^{'},b)}f_{a,t} - \sum_{a(b,b^{'})}f_{a,t} = D_{b,t} \ , \\
     & f_{a,t}^{min} \leq f_{a,t} \leq f_{a,t}^{max}, \\
     & LS_{b,t} \leq D_{b,t} + P_{crank, g_b} \ , \\
     & \forall \ {t \in T}, \ \forall \ {b \in B}, \ \forall a \in A_{nd},
    \end{split}
    \end{equation}
    where $f_{a,t}$ is the power flow along a non-damaged/restored line $a$ at step $t$; $A_{nd}$ is the set of non-damaged/restored lines; $D_{b,t}$ is the demand at bus $b$; $P_{g_b,t}$ is the power generated by generator $g$ at bus $b$, which has different limitations regards to BS and NBS units. 
    
When $g_b \in G_{BS}$,
    \begin{equation*}
        P_{g_b,t}^{min} \leq P_{g_b,t} \leq P_{g_b,t}^{max}.
    \end{equation*}
    
When $g_b \in G_{NBS}$,
    \begin{equation*}
        P_{g_b,t} = -P_{crank, g_b},
    \end{equation*}
    where $G_{BS}$ is the set of BS generators, and $G_{NBS}$ is the NBS set. $P_{crank, g_b}$ represents the cranking power needed to energize the NBS units, and the negative operation represents the NBS units are treated as loads before they are energized.
    
\subsection{Recovery Constraints}
During the restoration process, if a transmission line is recovered, the two buses connected will then be back to normal and the corresponding demand will also be supplied. Regards to the NBS buses, only when the lines connected to them are restored as well as the power flow equals to the cranking power, they will be energized and operate as dispatchable generators afterwards. Therefore, two types of constraints are proposed for the recovery process: line recovery and NBS recovery constraints.
    
\subsubsection{Line Recovery Constraints} With respect to the damaged lines, each line can only be recovered in a single recovery time period, meaning it will work normally after it's restored. 
    \begin{equation}
    \label{line}
    \begin{split}
        & B_{a,t} \in \{0, \ 1\}, \quad S_{a,t} = \sum_{t^{'}\leq t} B_{a,t^{'}}, \\
        & 0 \leq S_{a,t} \leq 1, \ \ \ \forall a \in A, \ \forall t \in T,
    \end{split}
    \end{equation}
    where $B_{a,t}$ is a binary variable that represents the status of each transmission line, with $1$ meaning the line is recovered or non-damaged, and $0$ meaning it's out of service. $S_{a,t}$ is a decision variable which is the integral over $B_{a,t}$: after a line is recovered ($B_{a,t}=1$), $S_{a,t}$ which represents the line status will keep to $1$ all the time periods afterwards.
    
    The power flow limits of damaged lines are,
    \begin{equation}
    \label{daline}
        S_{a,t} \cdot f_{a,t}^{min} \leq f_{a,t} \leq S_{a,t} \cdot f_{a,t}^{max}, \ \forall a \in A_d, \ \forall t \in T.
    \end{equation}
    Moreover, there's a budget limit on the amount of lines can be restored per time period,
    \begin{equation}
    \label{budget}
        \sum_{a \in A_{d}} B_{a,t} \leq L_{budget}, \ \forall t \in T,
    \end{equation}
    where $A_d$ is the set of damaged lines, and $L_{budget}$ is a parameter set in advance based on the recovery task requirements.
    
\subsubsection{NBS Recovery Constraints} 
\begin{figure}[htbp]
    \centering
  \centerline{\includegraphics[width=0.25\textwidth]{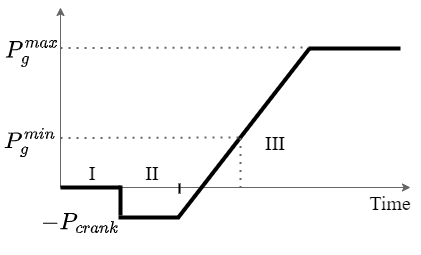}}
    \caption{NBS Generator Startup Curve.}
    \label{NBS}
\end{figure}

Fig. \ref{NBS} shows the startup curve of NBS generators which includes three phases. The NBS generator is out of service at phase I. At phase II, the NBS unit is energized after absorbing the cranking power and then ramp up to operate as a dispatchable generator (phase III). 
    Therefore, the status update of NBS units depends on the power flow to the buses linked them, making a coordination between the line recovery and NBS energization. 
    \begin{equation}
    \label{nbs}
    \begin{split}
        & \beta_{b,t} = \frac{\sum_{a(b^{'},b)}f_{a,t}-\sum_{a(b,b^{'})}f_{a,t}}{P_{crank, g_b}}, \\
        & \mu_{b,t} = \max (0, \ \beta_{b,1},\cdots, \ \beta_{b,t}), \\
        & \forall b \in G_{NBS}, \ \forall t \in T,
    \end{split}
    \end{equation}
    where $\beta_{b,t}$ is a variable that denotes the real-time NBS status, $\mu_{b,t}$ is a binary variable that sets the NBS status to $0$ (non-energized) or $1$ (energized).
    
    From \eqref{nbs}, it is clear that $\beta_{b,t}^{max}=1$, only when the power flow to an NBS unit $g_b \in G_{NBS}$ equals to the cranking power it needs, and $\mu_{b,t}$ will keep to $1$ afterwards.
    
    In addition, the power generation of an NBS unit is related to its status,
    \begin{equation}
    \label{nbsp}
    \begin{split}
        & -P_{crank, g_b} \cdot (1-\mu_{b,t}) + \mu_{b,t} \cdot P_{g_b}^{min} \ \leq \ P_{g_b,t} \\
        & \leq -P_{crank, g_b} \cdot (1-\mu_{b,t}) + \mu_{b,t} \cdot P_{g_b}^{max}, \ g_b \in G_{NBS}. 
    \end{split}
    \end{equation}
\eqref{nbsp} indicates that before the NBS unit is energized, it generates negative power which means it operates as a load, and then it will work as a dispatchable generator after getting the related cranking power (when $\mu_{b,t}=1$).

Summarized from the historical dataset in \cite{cranking}, the amount of cranking power required by NBS resources is set to be,
\begin{equation}
\label{crank}
    P_{crank, g_b} = 0.1 \cdot P_{g_b}^{max}, \ \forall g_b\in G_{NBS}.
\end{equation}

\section{NBS Recovery}
The previous Section introduces the mathematical model of the recovery task. Among the constraints, \eqref{nbs} represents a maximization function which needs to be reformulated to retain a MILP model for the optimization problem. Solving the optimization problem subject to a $max$ function is NP-hard. Therefore, the $max$ function in \eqref{nbs} should be linearized first by using linear programming. Inspired by \cite{3568461}, the $max$ function in \eqref{nbs} is replaced with a linear formulation,
\begin{equation}
\label{max}
    \begin{split}
       & \mu_{b,t} \ \geq 0, \ \mu_{b,t} \ \geq \beta_{b,i} \\
       & \mu_{b,t} \leq \beta_{b,i} + (1-\epsilon_i)\times M \\
       & \sum_{i \in t}\epsilon_i = 1,
    \end{split}
\end{equation}
where $\epsilon_i$ is a binary variable that when $\epsilon_i=1$, $\beta_{b,i}$ is the maximum value; $M$ is a big number which is set in advance.

As a result, the objective function \eqref{obj} and the constraints \eqref{ope}-\eqref{max} constitute  the complete MILP recovery optimization model.

\section{Simulation Results}
The proposed MILP recovery model is validated on several test systems: $9$-bus \cite{2013LocalSO}, $39$-bus \cite{2013LocalSO}, $113$-bus systems, and a realistic representation of Puerto Rico (PR) transmission network. The model is written in GAMS and solved with CPLEX. These tests are conducted on a laptop with Intel Core i5 ($2.5$GHz) and an $8$GB RAM.

\subsection{Case Study 1: 9-bus system}
In the $9$-bus system, buses $1$-$3$ are generators, where bus $1$ has the only BS unit. There are $9$ transmission lines in the system and no more than $3$ lines can be restored at each step. 

\begin{figure}[htbp]
    \centering
  \begin{subfigure}[b]{0.22\textwidth}
    \includegraphics[width=\textwidth]{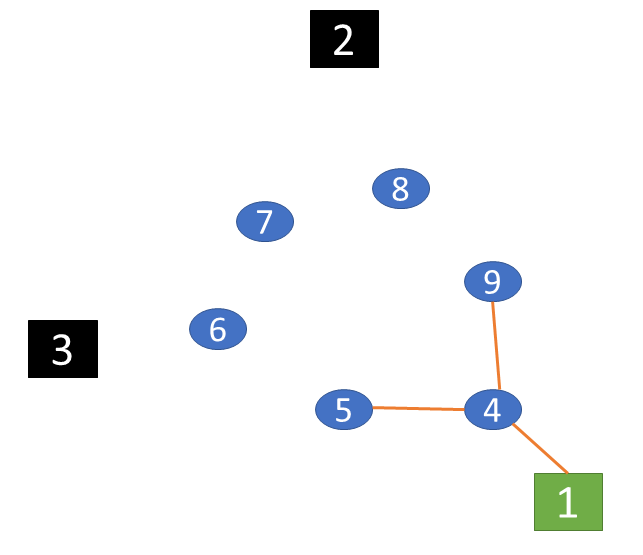}
    \caption{Step 1}
  \end{subfigure}
  \begin{subfigure}[b]{0.22\textwidth}
    \includegraphics[width=\textwidth]{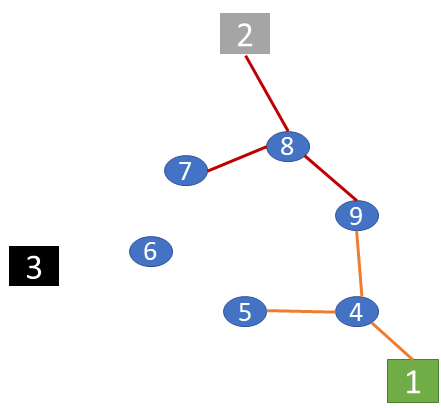}
    \caption{Step 2}
  \end{subfigure}
  \begin{subfigure}[b]{0.22\textwidth}
    \includegraphics[width=\textwidth]{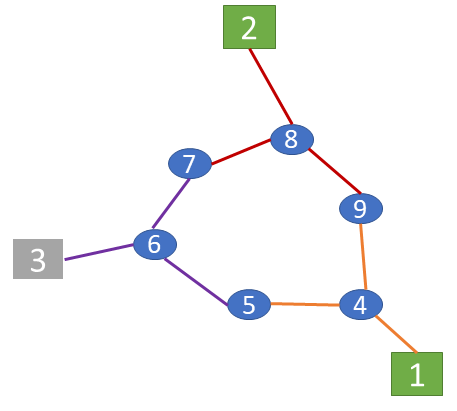}
    \caption{Step 3}
  \end{subfigure}
  \begin{subfigure}[b]{0.22\textwidth}
    \includegraphics[width=\textwidth]{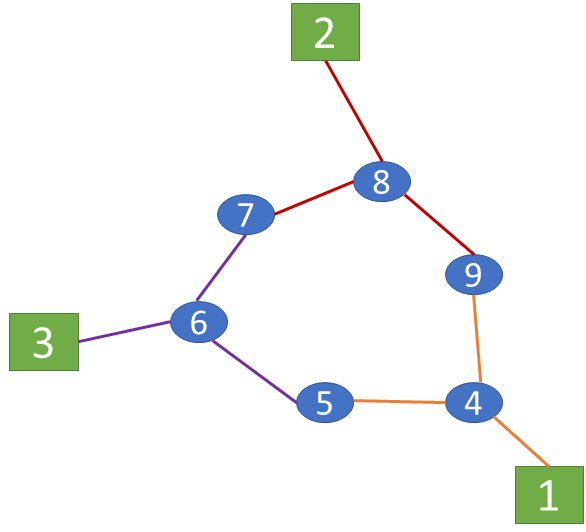}
    \caption{Step 4}
  \end{subfigure}
    \caption{Recovery Process of 9-bus Power System (blue ellipse with bus number: loads; green rectangular with bus number: BS unit or restored NBS; black rectangular with bus number: NBS unit; grey rectangular with bus number: NBS unit is energized; colorful lines: recovered transmission lines denoted by different colors at different steps).}
    \label{9bus}
\end{figure}

\begin{figure*}[htbp]
    \centering
  \begin{subfigure}[b]{0.3\textwidth}
    \includegraphics[width=\textwidth]{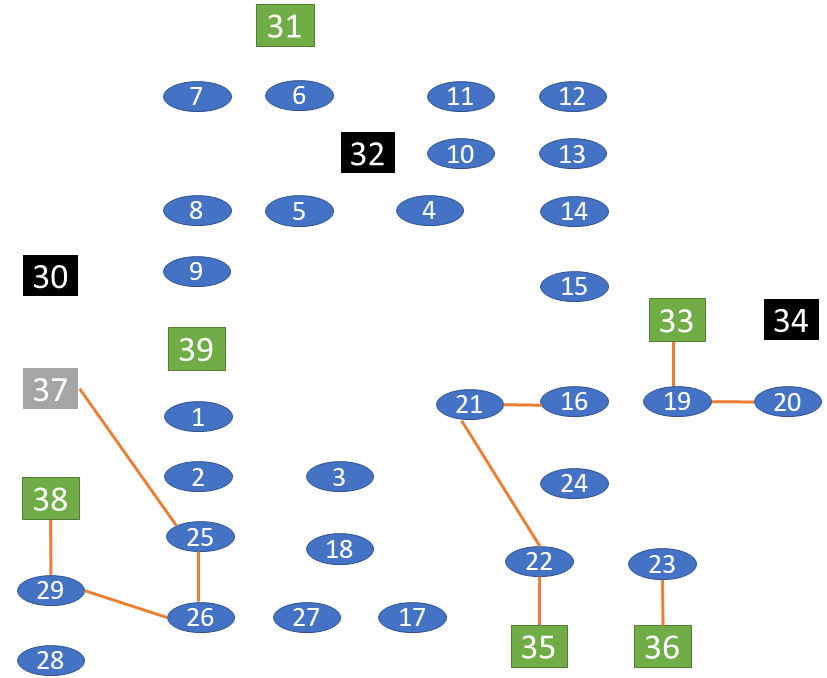}
    \caption{Step 1}
  \end{subfigure}
  \begin{subfigure}[b]{0.3\textwidth}
    \includegraphics[width=\textwidth]{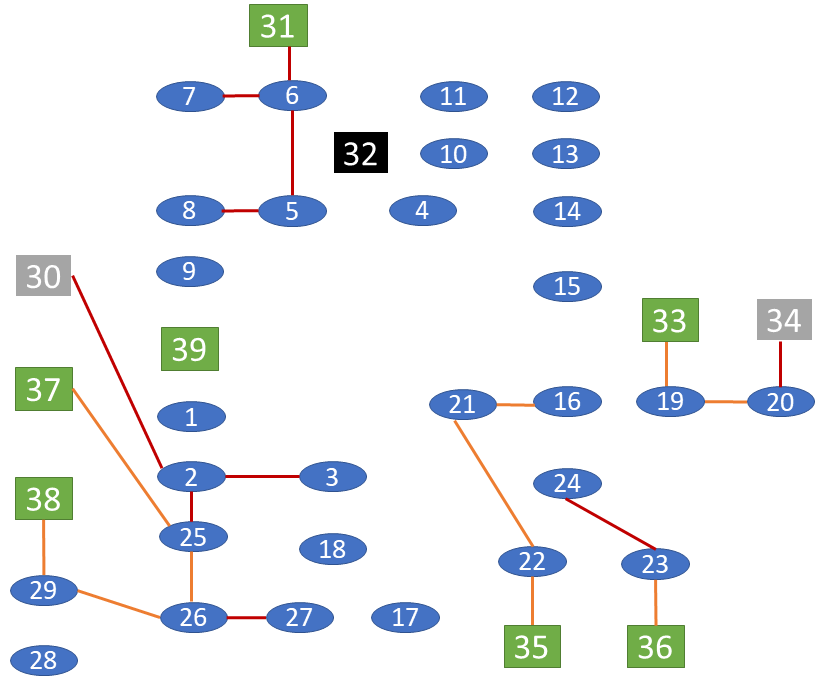}
    \caption{Step 2}
  \end{subfigure}
  \begin{subfigure}[b]{0.3\textwidth}
    \includegraphics[width=\textwidth]{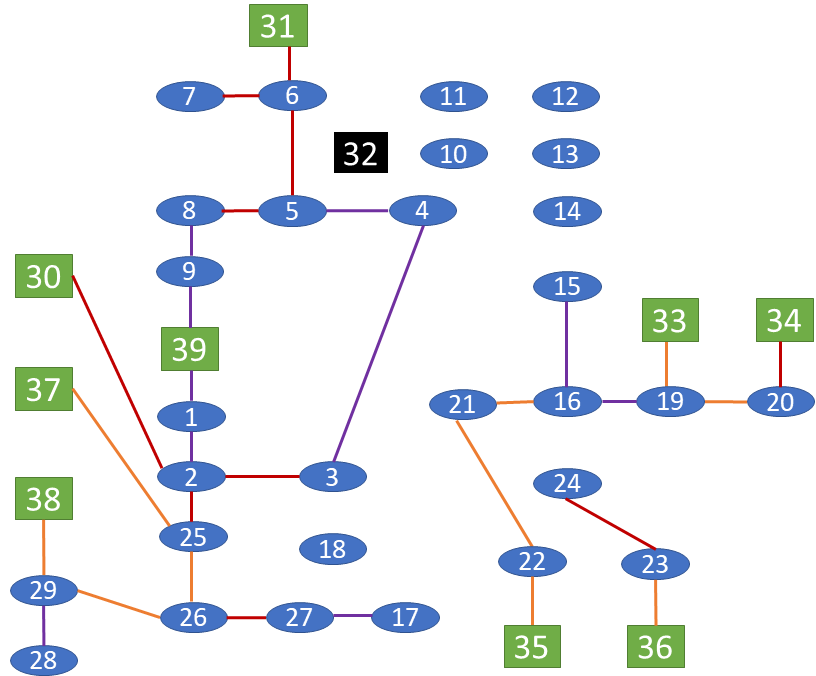}
    \caption{Step 3}
  \end{subfigure}
  \begin{subfigure}[b]{0.3\textwidth}
    \includegraphics[width=\textwidth]{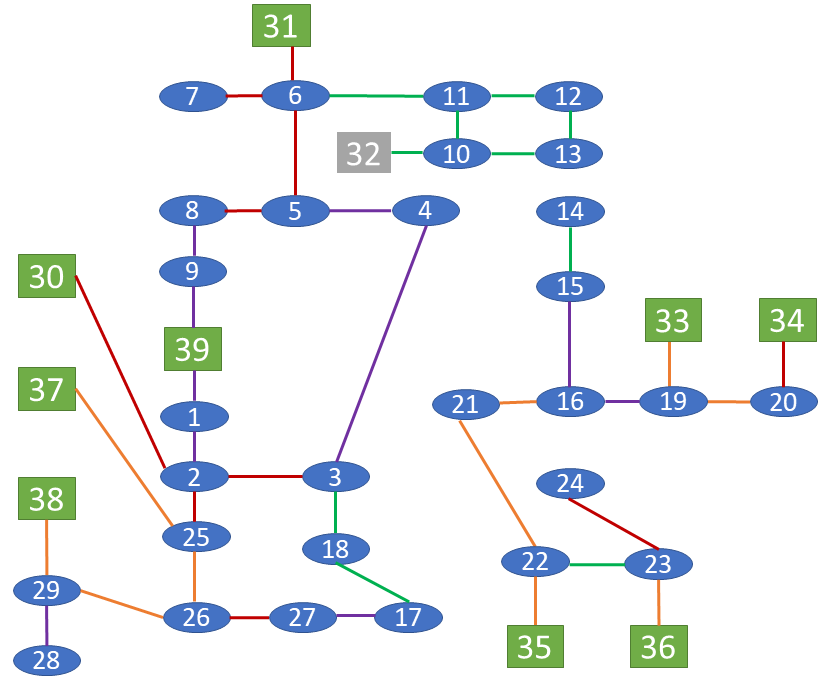}
    \caption{Step 4}
  \end{subfigure}
  \begin{subfigure}[b]{0.3\textwidth}
    \includegraphics[width=\textwidth]{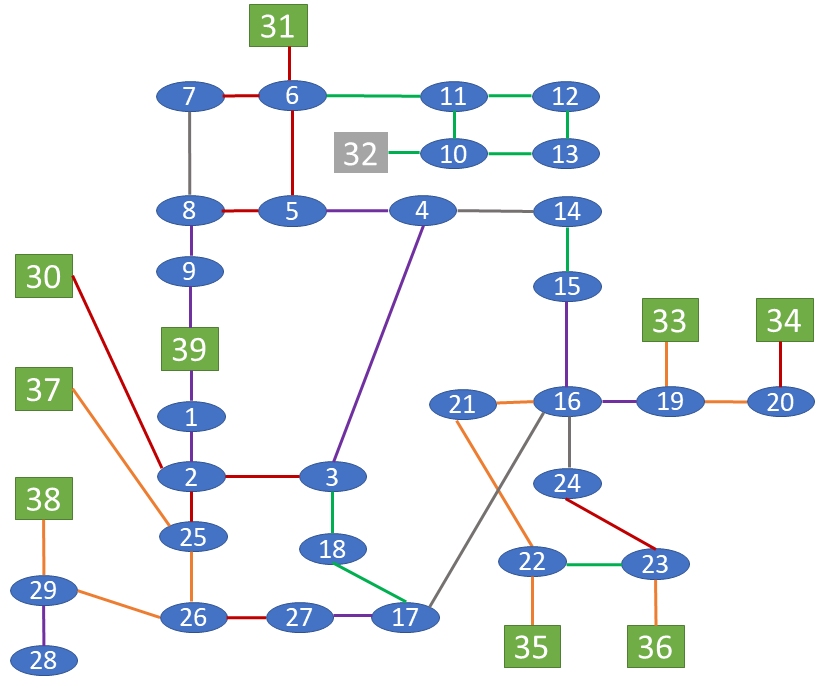}
    \caption{Step 5}
  \end{subfigure}
  \begin{subfigure}[b]{0.3\textwidth}
    \includegraphics[width=\textwidth]{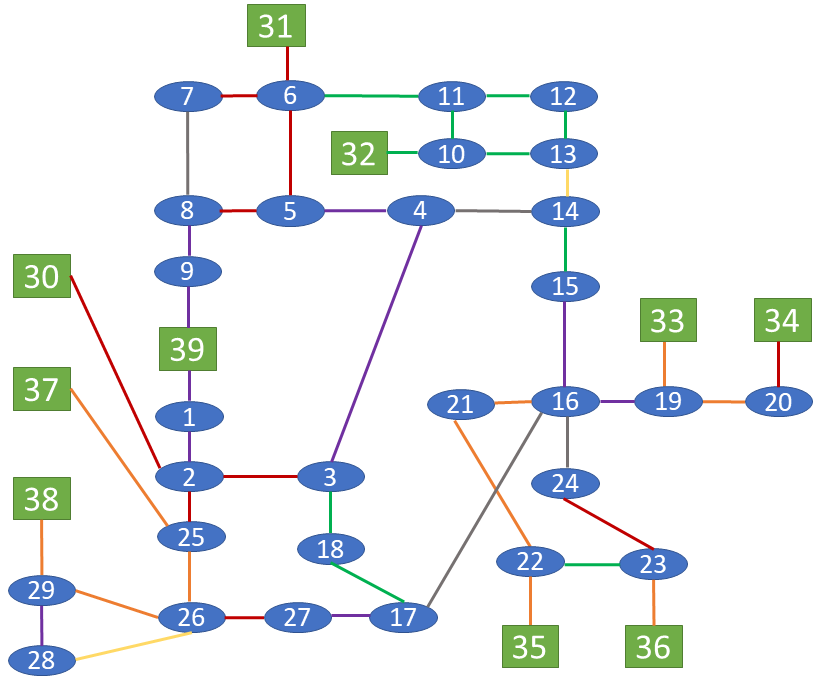}
    \caption{Step 6}
  \end{subfigure}
    \caption{Recovery Process of 39-bus Power System (blue ellipse with bus number: loads; green rectangular with bus number: BS unit or restored NBS; black rectangular with bus number: NBS unit; grey rectangular with bus number: NBS unit is energized; colorful lines: recovered transmission lines denoted by different colors at different steps).}
    \label{39bus}
\end{figure*}

The recovery process of this $9$-bus system is shown in Fig. \ref{9bus}. It is notable that the loads connected with BS generators were recovered first due to unserved load minimization as well as the recovery budget limitation; the NBS buses linked with the recovered buses were energized afterwards and then operated as dispatchable generators to supply the system.


\subsection{Case Study 2: 39-bus system}
In the $39$-bus system, there are $10$ generators located at buses $30$-$39$ in which buses $30$, $32$, $34$, $37$ are allocated with NBS resources, $39$ loads and $46$ transmission lines.

\begin{figure}
    \centering
  \centerline{\includegraphics[width=0.35\textwidth]{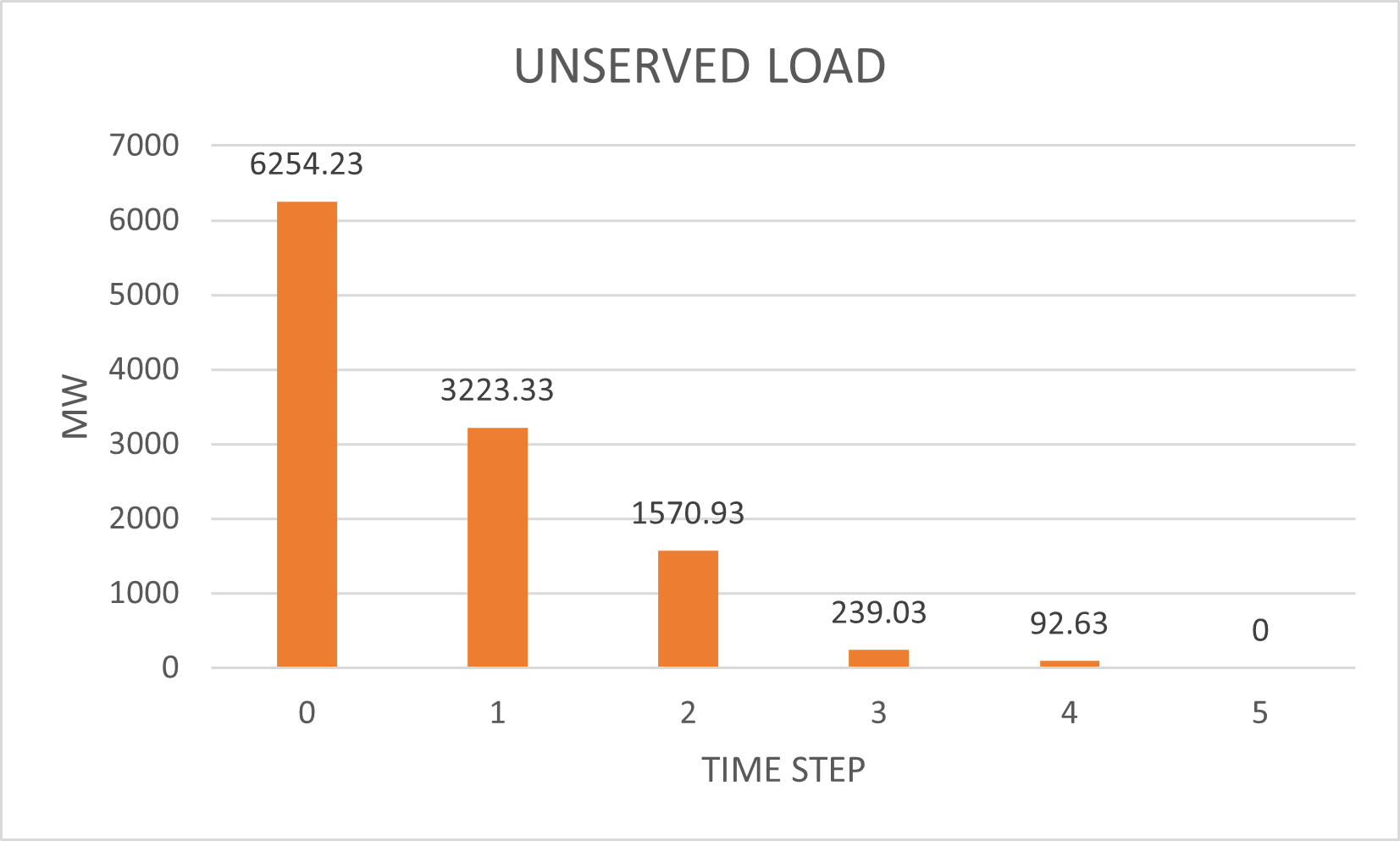}}
    \caption{Unserved Load Trend of 39-bus System.}
    \label{39load}
\end{figure}

Fig. \ref{39bus} shows the recovery process of the $39$-bus system, and Fig. \ref{39load} illustrates the total system unserved loads during the recovery period. The restoration budget is $10$ transmission lines at each recovery step. At time step $1$, the critical loads were recovered first (half of the loads were recovered as is shown in Fig. \ref{39load}), and some NBS units were energized at step $2$. It is obvious that the NBS recovery and transmission lines restoration are conducted in parallel, resulting in a well-coordinated recovery process. Since some NBS resources were energized at earlier steps, the unserved loads reduced dramatically and all of the loads were recovered at step $5$.

\subsection{Case Study 3: 113-bus system}
In the $113$-bus system, $29$ generators are installed with $10$ of them are NBS units, the number of loads and transmission lines are $69$ and $149$, respectively. 

\begin{figure}
    \centering
  \begin{subfigure}[b]{0.35\textwidth}
    \includegraphics[width=\textwidth]{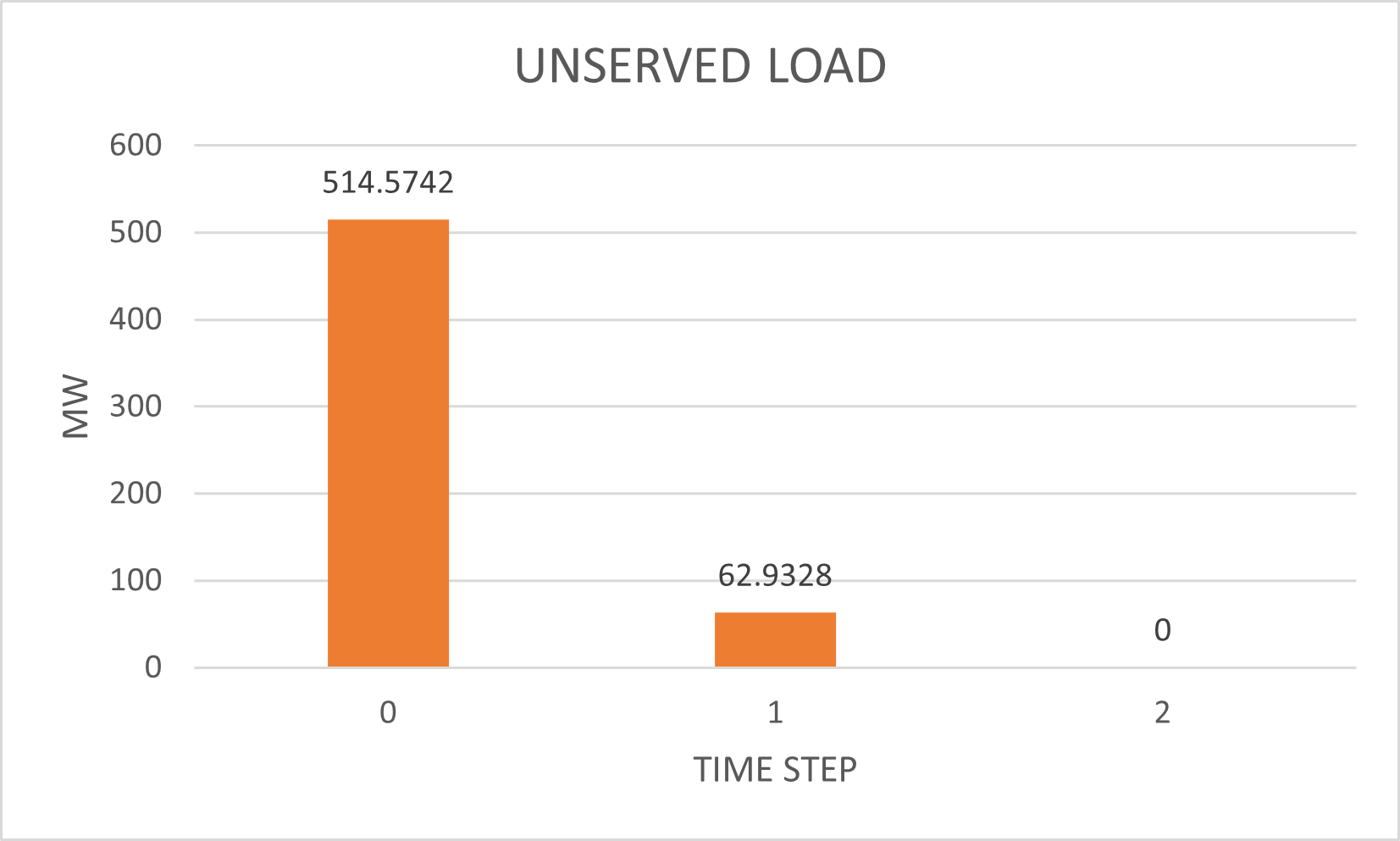}
    \caption{Unserved Load Trend along with Restoration Steps}
  \end{subfigure}
  \begin{subfigure}[b]{0.35\textwidth}
    \includegraphics[width=\textwidth]{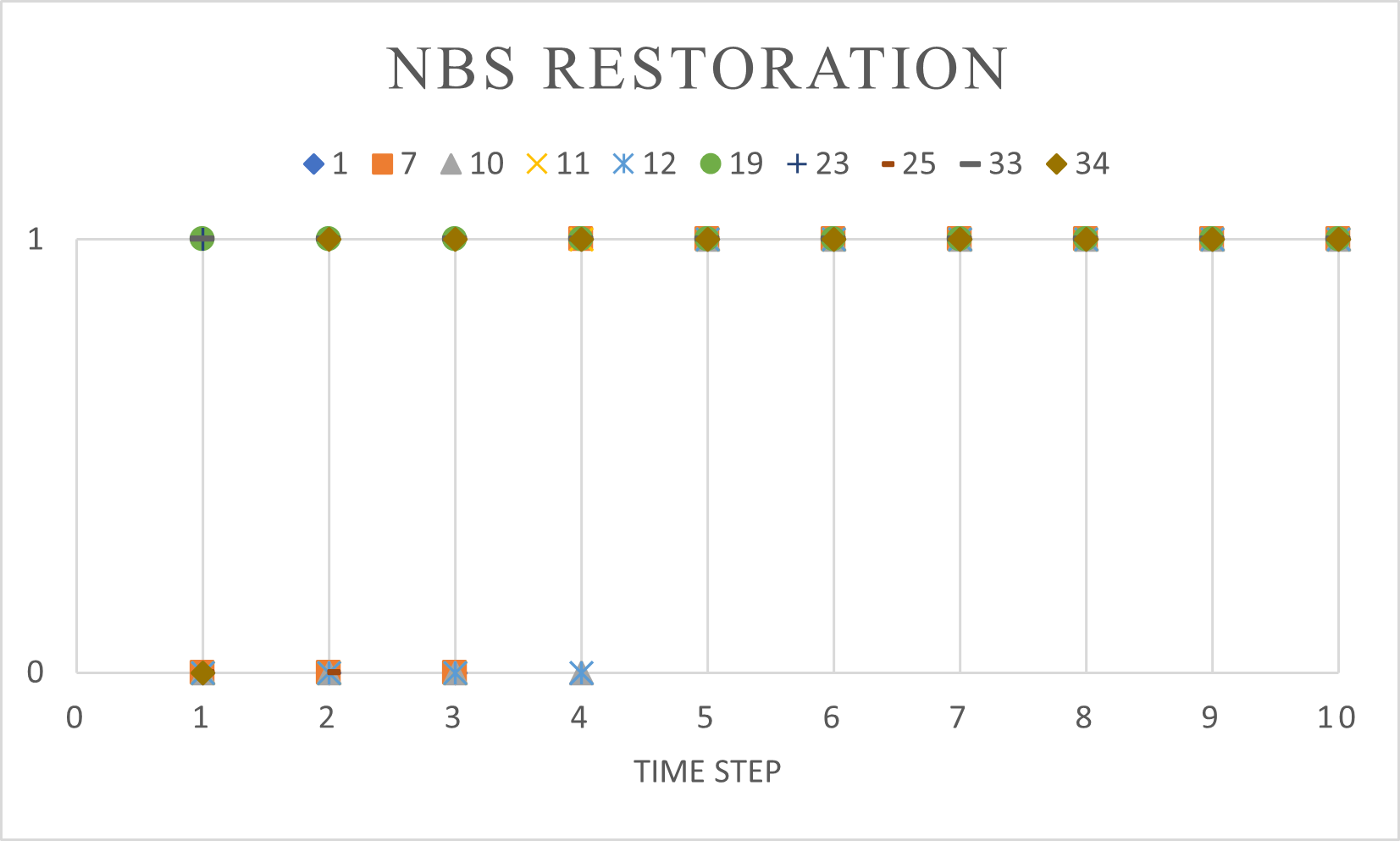}
    \caption{NBS Recovery along with Restoration Steps}
  \end{subfigure}
    \caption{Recovery Summation of 113-bus Power System.}
    \label{113bus}
\end{figure}

During the recovery process, the restore budget is set to $30$ and the number of time steps is $10$. Fig. \ref{113bus} summarizes the recovery process with the unserved load trend and NBS restoration at each time step. It shows that most of the loads were recovered at time step $1$, as well as some NBS units. At step $2$, all of the loads were recovered, and all of the NBS resources were restored at step $5$, demonstrating an efficient and well-coordinated recovery process.

\subsection{Case Study 4: Puerto Rican (PR) $1393$-bus system}
In the PR $1393$-bus system \footnote{While the transmission system is a representative of the PR system, the outage data is not from a real event but rather configured by the authors. Similarly, units modeled as BS/NBS are configured by the authors to determine the scalability of the model.}, $113$ generators are allocated with $22$ of them are NBS units; $823$ load buses; $1549$ transmission lines and about $88$\% of them are damaged. 

\begin{figure}
    \centering
  \begin{subfigure}[b]{0.35\textwidth}
    \includegraphics[width=\textwidth]{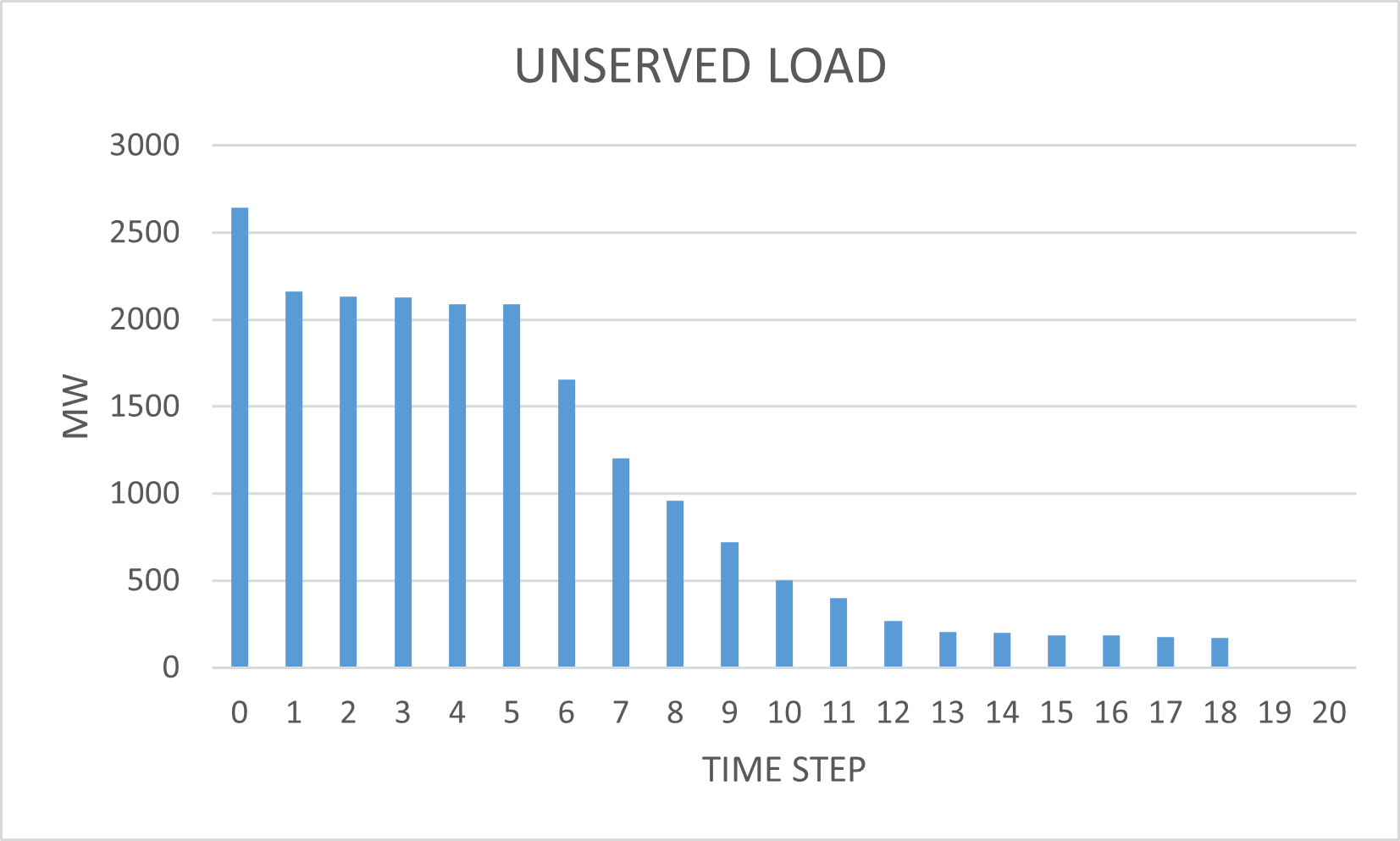}
    \caption{Unserved Load Trend along with Restoration Steps}
  \end{subfigure}
  \begin{subfigure}[b]{0.35\textwidth}
    \includegraphics[width=\textwidth]{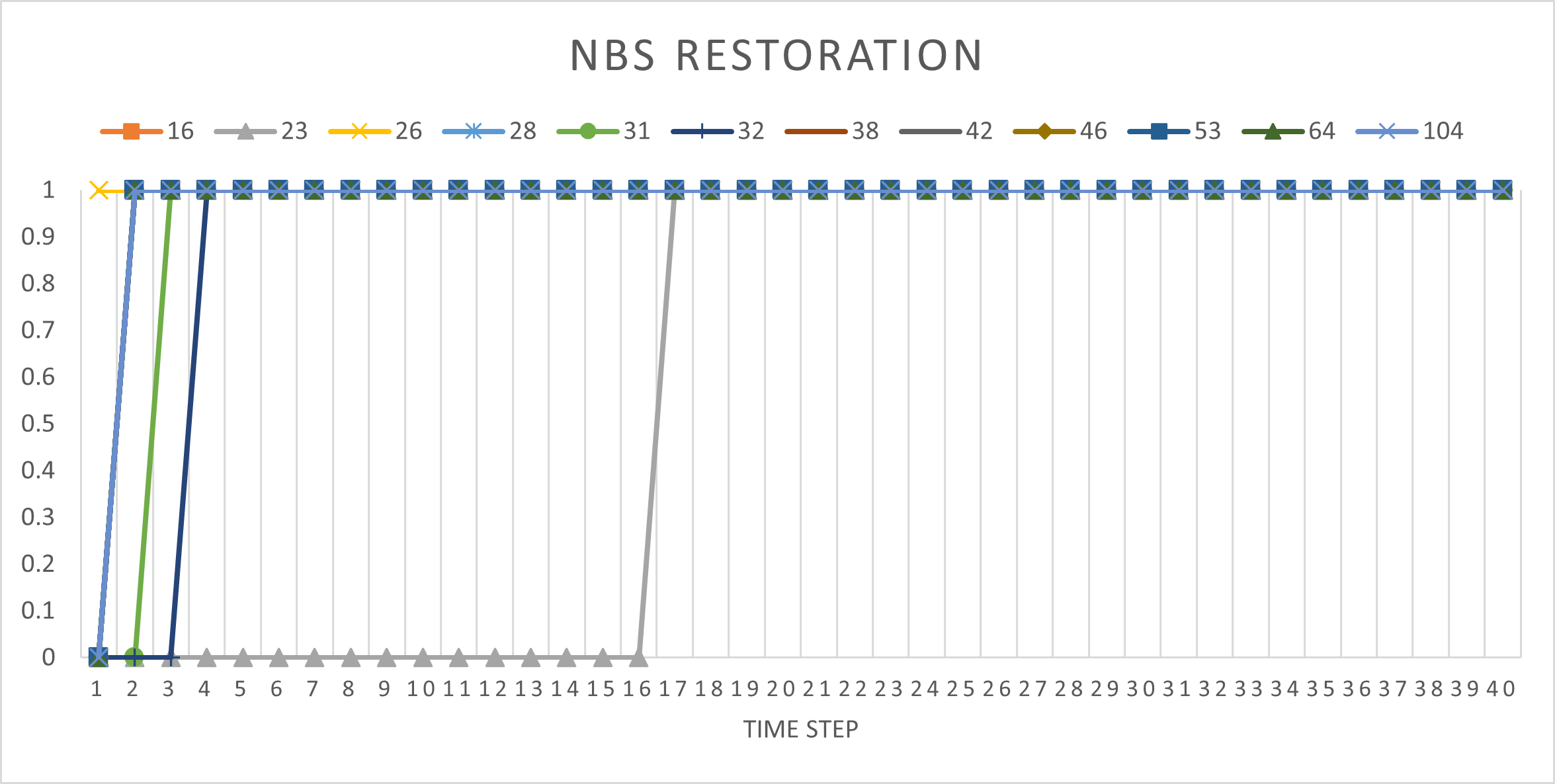}
    \caption{NBS Recovery along with Restoration Steps}
  \end{subfigure}
    \caption{Recovery Summation of PR 1393-bus Power System.}
    \label{real}
\end{figure}

During the restoration process, no more than $100$ transmission lines can be recovered. The restoration summation which includes unserved load trend and NBS recovery is shown in Fig. \ref{real}. 
It is clear that most of the NBS units were restored before step $3$, and the loads were all served at step $19$. However, the unserved load trend changed slowly from step $2$-$5$. This is likely because only half of the buses need to be served and most of them are not connected closely with the generators. Therefore, to supply power to the demand buses, the intermediate transmission lines should be recovered first even though there are no load buses. After step $5$, the unserved loads decreased dramatically at each recovery step since the recovery paths between the generator and load buses were well established by the intermediate lines.

\section{Conclusion}

With regard to the system recovery after power blackouts, this paper proposes a MILP optimization model by solving a direct-current optimal power flow (DCOPF) problem. This model considers the NBS restoration as well as the transmission line recovery simultaneously, and their status are denoted with binary variables. Since cranking power is needed to energize the NBS units, the NBS resources are regarded as loads and their status update is coordinated with the power flow to them. Based on historical data, the amount of cranking power is approximated as a specific percentage of the maximum generation capacity. Simulation results demonstrate that the NBS restoration and transmission line recovery coordinate well during the recovery process, resulting in a more efficient and optimal restoration model.

Limitations of current model include: each bus deployed an NBS unit cannot have additional generators, which is not ideal for real-world systems. Therefore, overcoming this limitation would be the focus of future work, as well as increasing the computation speed of the algorithm to enhance the recovery formulation's application to real-world problems.




\bibliographystyle{IEEEtran}
\bibliography{reference}

\end{document}